\documentclass[journal,10pt,twocolumn]{IEEEtran}

\usepackage{cite}

\usepackage{amsthm}

\usepackage{graphicx}

\usepackage{verbatim}

\usepackage{psfrag}

\usepackage{subfigure}
\usepackage{amsmath}
\usepackage{epstopdf}
\usepackage{algorithmic}

\newtheorem{definition}{Definition}
\newtheorem{example}{Example}

\usepackage[longend,ruled]{algorithm2e}

\usepackage{booktabs}

\newcommand{\mat}{\boldsymbol}

\newcommand{\AND}{\mathbf{AND}}

\begin{document}
\title{Energy and Bursty Packet Loss Tradeoff over Fading Channels: A System Level Model}

\author{M.~Majid~Butt,~\IEEEmembership{Senior~Member,~IEEE,}
Eduard A. Jorswieck,~\IEEEmembership{Senior~Member,~IEEE,} and\\
Amr Mohamed,~\IEEEmembership{Senior~Member,~IEEE}

\thanks{This work was made possible by NPRP grant 5-782-2-322 from the Qatar National Research Fund (a member
of Qatar Foundation). The statements made herein are solely the responsibility of the authors.}

\thanks{
M. Majid Butt was with Computer Science and Engineering Department, Qatar University, Doha, Qatar. He is now
with CONNECT center for future networks, Trinity college Dublin, Dunlop Oriel House, Dublin 2, Ireland.
E-mail: majid.butt@ieee.org.}
\thanks{Eduard A. Jorswieck is with Department of Electrical Engineering and Information Technology, TU
Dresden, Dresden 01062, Germany. Email: eduard.jorswieck@tu-dresden.de.
}
\thanks{Amr Mohamed is with Computer Science and Engineering Department, Qatar University, Doha, Qatar.
E-mail: amrm@qu.edu.qa.}

}

\maketitle

\begin{abstract}
Energy efficiency and quality of service (QoS) guarantees are the key design goals for the 5G wireless
communication systems. In this context, we discuss a multiuser scheduling scheme over fading channels for
loss tolerant applications. The loss tolerance of the application is characterized in terms of different
parameters that contribute to quality of experience for the application. The mobile users are scheduled
opportunistically such that a minimum QoS is guaranteed. We propose an opportunistic scheduling scheme and
address the cross layer design framework when channel state information is not perfectly available at the
transmitter and the receiver. We characterize the system energy as a function of different QoS and channel
state estimation error parameters. The optimization problem is formulated using Markov chain framework and
solved using stochastic optimization techniques. The results demonstrate that the parameters characterizing
the packet loss are tightly coupled and relaxation of one parameter does not benefit the system much if the
other constraints are tight. We evaluate the energy-performance trade-off numerically and show the effect
of channel uncertainty on the packet scheduler design.
\end{abstract}


\section{Introduction}
\label{intro}
Energy efficient (green) communication is one of the design principles for the next generation of wireless
networks. Energy efficiency in a network can be defined in terms of different metrics, e.g., bits-per-Joule,
which takes network load and static circuit consumed energy into account \cite{Feng:comsurveys_2013}. In this
work, we employ signal-to-noise-ratio (SNR) per information bit, $E_b/N_0$ as a measure of energy efficiency
to focus more on communication aspect of the problem. Due to high electricity cost of operating a network,
the revenue generation for the network operators is vanishing and the network architecture design requires a
complete new design methodology.
quality of service (QoS) metrics.
Energy efficiency can be achieved by trading bandwidth, delay or other system performance indicators
\cite{Chen:11}. At the same time, energy efficiency can be achieved by architecture level novel techniques
which include switching off the base stations, cell breathing and sleep mode design \cite{Wu:2013,Niu:2011}.

The QoS for a service is measured by the parameters like throughput, delay and packet loss tolerance. These
parameters control the quality of experience (QoE) for the end user. For example, if the application is delay
sensitive, a large amount of radio resources are required to meet the QoS requirements as compared to delay
tolerant applications. Depending on the network design constraints, these resources result in either large
bandwidth or excessive use of power. As allocated bandwidth for a system is fixed usually, it is important to
exploit any relaxation in QoS parameters to make the system more energy efficient.

This work aims at exploiting such relaxed application QoE requirements to achieve system energy efficiency.
In literature, energy--delay tradeoffs have been addressed in different settings, e.g.,
\cite{Todd,Neely,Zhang:2013}. However, not much work focuses on exploiting the loss tolerance of the
application in radio resource allocation mechanisms due to the challenging task of providing a certain
guaranteed QoE. By a service provider's point of view, if a user's application can tolerate a certain amount
of data loss without deteriorating QoE significantly, it is advantageous to exploit it for overall system
efficiency. By the end user's point of view, it is not really advantageous to pay for an extra quality when
it is not needed. The application's loss tolerance acts as a degree of freedom (DoF) that can be exploited to
make system energy efficient. The dynamically fading wireless channel poses an interesting challenge of
scheduling the packets optimally such that QoE for the end user remains acceptable (bounded QoS) while the
extra packets are intentionally dropped at the transmitter to save transmission energy. It should be noted
that random packet dropping with average packet drop rate guarantee cannot promise required QoE as there are
additional QoS key factors involved in perception. For example, bursty packet loss causes fast degradation in
QoE as compared to some random packet loss pattern even for the case when the average packet loss remains the
same. In addition to average packet drop rate, we consider bursty nature of the packet loss as a constraint
on packet scheduling and analyze our scheme such that a minimum (promised) QoS is provided and the system
energy efficiency is improved at the same time.

\subsection{Related Works}
In literature, packet loss or packet dropping mechanisms are usually treated as higher layer issues. Though,
a lot of work models and analyzes the effect of packet dropping on QoS, most of the work focuses on
traditional wired networks or protocol level mechanisms without taking unpredictable wireless channel into
account. In \cite{Miyata:2001} successive packet loss modeling is considered using Markov chain analysis.
The work in \cite{Altman:2000} investigates the sensitivity of the time average of the transmission rate on
the distribution of losses and the average loss rate for the flow control using transmission control
protocol. The authors show  that the time average of the transmission rate increases with the burstiness of
losses for a given average packet loss rate.
The authors in \cite{Sequeira:2013} present an analysis for the effect of the access router buffer size on
packet loss rate and determine its effect on the QoS of multimedia services when bursty traffic is present.
The study shows that the bursty nature of some applications impairs multimedia traffic especially when a
certain number of bursts overlap.
Fanqqin \emph{et al.} discuss a useful analytical framework to dimension the packet loss burstiness over
generic wireless channels \cite{Fanqqin:2013}. They propose a new metric to characterize the packet loss
burstiness, which is shown to be more compact and accurate than the metrics proposed previously.
packet loss performance in terms of packet loss rate and the loss-burst length distributions.
flows.

It is apparent that successive or bursty packet loss has been investigated quite a bit in the past, but this
dimension has not been the focus of much research in the wireless domain. The works in
\cite{Neely2006,Neely2009} consider intentional packet dropping mechanisms for delay limited systems to
minimize energy consumption over fading channels. Some recent works in \cite{majid_TWC:13,majid_TWC:15}
consider data loss tolerance as an other aspect of the system which can be exploited to save system energy.
The authors in \cite{majid_TWC:13} introduce a framework to achieve energy efficiency in a multiuser multiple
access system for an application with average packet loss and maximum successive packet loss constraint. This
work is generalized to a system with a finite buffer size in \cite{majid_TWC:15} and it analyzes the bounds
on buffer size for the loss tolerance parameters.

\subsection{Contributions and Main Results}
The works in references \cite{majid_TWC:13,majid_TWC:15} consider perfect channel state information (CSI) at
both transmitter and receiver sides. The sequence of maximum number of packets allowed to be dropped
successively for a given average packet drop rate $\theta_{\rm tar}$ is termed as \emph{continuity
constraint} (CCON) parameter and denoted by $N$. Every user of the application is provided a guaranteed QoS
in terms of metrics $(N,\theta_{\rm tar})$ with probability one while the CCON parameter is identical for all
users.

This work extends the work in \cite{majid_TWC:15} to the cases when CSI available at the transmitter and the
receiver is not perfect, which logically translates into the problem of providing statistical guarantees on
$N$ to the individual users.

For our problem settings, we have two reasons for a packet drop:
\begin{enumerate}
  \item {Intentional packet drop at the transmitter depending on the application loss tolerance to save
      energy if the applications's loss tolerance
      permits.}
  \item {Packet drop due to imperfect CSI estimate at the transmitter (and receiver) side which implies
      that the actual channel state is worse than the estimated one and results in
      packet loss after
      transmission.}
\end{enumerate}
The energy efficient scheduling algorithm design for the packet loss tolerant applications takes the packet
loss due to imperfect CSI into account statistically and adapts its intentional packet drop rate accordingly
to maintain a bound on $\theta_{\rm tar}$ and $N$ parameters.

The main contributions of this paper are summarized as follows:
\begin{itemize}
  \item We use a packet level channel model to model the effect of imperfect CSI on the transmitter side
      and analyze the proposed scheduling scheme as a function of different parameters that govern the
      QoS.
  \item We generalize the framework to the case when the individual users have their own CCON parameters
      and model it at system level as a Markov decision process. The system energy depends on the
      distribution of the CCON parameter.
  \item Then, the proposed scheme is analyzed when the CSI estimation error at both transmitter and
      receiver sides is modeled by error variance. The energy per transmitted bit is derived in closed form
      for a multiuser multiple access system as a function of error variance.
techniques to solve the optimization problem for an energy efficient system.
  \item The loss tolerance for the application's QoE is controlled by different parameters as we discussed.
      We study the \emph{coupling} effects of these parameters on the system energy through simulations.
      The coupling effect implies that a tight requirement on one of the loss parameters implies that there
      is a bound on the maximum exploitation of the other parameters as well, and further energy efficiency
      cannot be achieved by relaxing the other parameters.
\end{itemize}

The rest of this paper is organized as follows. Section \ref{sec:system_model} introduces the system model
and key assumptions used in the analysis. We model the proposed scheduling scheme in Section
\ref{sect:scheme}. The optimization problem is formulated mathematically in Section \ref{sect:optimization}
and Section \ref{sect:generalization} addresses the generalization of the framework. The tradeoff between
energy and QoS parameters is evaluated numerically in Section \ref{sect:results} and Section
\ref{sect:conclusions} concludes with the main contributions of the work.

\section{System Model}
\label{sec:system_model}
We assume that $K$ users in a multiple access channel (MAC) are uniformly and randomly distributed in a
wireless network with a base station in the center. The user scheduled in a time slot is provided an average
rate $R_k=\lambda_k\frac{C}{K}$ where $C$ is the system spectral efficiency and $\lambda_k$ denotes a random
variable \cite{Ralf1,majid_TWC:13}.
\subsection{Propagation Channel Model}
We consider an uplink scenario where time is slotted such that each user $k$ experiences a channel gain
$h_k(t)$ in a time slot $t$.
Signal propagation is characterized by a distance dependent
path loss factor and a frequency-selective short-term fading.

Thus, $h_k(t)$ turns out to be
\begin{equation}
h_k(t)=s_kf_k(t)
\end{equation}
where $s_k$ and $f_k(t)$ denote the path loss and the short term fading of user $k$, respectively.

The users are assumed to be uniformly distributed
in a geographical area but for a forbidden circular region of radius
$\delta$ centered around the base station where $0<\delta\leq 1$ is
a fixed system constant \cite{Ralf1}. Using this model, the cumulative distribution function (cdf) of path
loss is
given by
\begin{equation}
{\rm F}_s(x) = \left\{
\begin{array}{c@{\qquad \qquad}c}
0& x < 1 \\
1-\frac{x^{-2/\alpha}-\delta^2}{1-\delta^2} &  1\leq x < \delta^{-\alpha}\\
1&x\geq \delta^{-\alpha}
\end{array}
\right.
\label{eqn:path_loss}.
\end{equation}
where the path loss at the cell border is normalized to one.
The path loss is assumed to be constant at the time scale considered in this work.
We assume block fading model such that the fading remains constant during a single time slot, but changes
with time slot. The fading is independently and identically distributed (i.i.d.) across both users and time
slots.

Thus, the multiple access channel (MAC) is described by input $X$ and output $Y$ relation by
\begin{equation}
Y_k(t)=\sum_{k=1}^K \sqrt{h_k(t)}X_k(t)+ Z(t)
\end{equation}
where $Z$ represents additive i.i.d. complex Gaussian
random variable with zero mean and unit variance.

This work focuses on leveraging the analysis of the scheme proposed in \cite{majid_TWC:15} when perfect CSI
is not available. Assumption of perfect CSI helps to perform system analysis and get insights about different
trade-offs involved in system design. However, acquisition of CSI is a costly operation and imperfection in
CSI causes performance degradation.

In this work, we consider two cases: \begin{itemize}
  \item Imperfect CSI at the transmitter side.
  \item Imperfect CSI at both transmitter and receiver sides.
\end{itemize}
The receiver acquires CSI using pilot or data aided channel estimation while acquisition of CSI at the
transmitter side requires feedback from the receiver. Feeding back information to the transmitter requires
transmission of a lot of side information and has an associated overhead cost. Specially, availability of CSI
at the transmitter side in a fast mobility scenario is very complex and the cost is enormous. This leads to a
tradeoff between \emph{exploration} and \emph{exploitation} \cite{Kaelbling:96,Chaporkar:2009}.

We consider different frameworks to analyze the effect of imperfection in CSI. We employ a simplified
framework for the case of imperfect CSI at the transmitter side (only). We model it using a packet level
channel model and adapt our scheduling decisions accordingly. When CSI is not available both at the
transmitter and the receiver sides, we model it by a channel estimation error variance and compute the
resulting energy per bit as a function of error variance.
\subsection{Packet Level Channel Model}
\label{sect:packet_level}
We assume that CSI is available at the transmitter side, but it is not perfect. Instead of modeling
imperfection statistically, we model it at packet level. As a result of imperfect CSI, the scheduled users
are not able to compute the correct power level for the assigned rate which could result in a packet loss. We
model this by a probability $\nu_d$ that a transmission is not successful. Furthermore, we assume that if the
transmission is not successful, all the scheduled packets are lost. The information about packet dropping is
fed back by the receiver to the transmitter by the end of time slot via a perfect channel. This model is
termed as packet level channel model in literature and has been investigated in different settings, e.g.,
\cite{Wei:2010,chie:2010}. As the one bit \emph{delayed} feedback information about the
successful/unsuccessful transmission of the previous packet arrives by the scheduling instance in the next
time slot, the transmitted packet(s) is buffered by then. If the transmission is successful, it is dropped
otherwise, it is taken into account for the scheduling decision in the next time slot depending on the buffer
capacity as explained later.
\subsection{Statistical Guarantees on Continuity Constraint}
The model considered in \cite{majid_TWC:15} assumes that CCON can be met with probability one. It is not
practicable to assume
that a packet can be transmitted with probability one over fading channels when $N$ packets have been dropped
successively. We generalize this framework in the direction of providing
statistical guarantees on CCON, i.e., a user violates the CCON with a probability $\gamma$. If channel
conditions are not good after dropping $N$ packets successively, the user is still allowed to drop a finite
amount of packets corresponding to $\gamma\geq 0$. We define the event of violation of CCON as the number of
time slots the packets are dropped after successively dropping $N$ packets.

We allow multiple users to be scheduled in a single time slot to minimize $\gamma$. If only a single user is
scheduled per time slot, all the users other than the scheduled one may have to drop the packets
(intentionally) which results in increase in $\gamma$ rapidly.
We have no control over the packets dropped due to channel impairments, but the packet scheduler can be
designed such that
$\gamma$ is bounded by facilitating maximum scheduling of the users who already have dropped $N$ packets
successively.

The analysis of the scheme is based on asymptotic user case which implies that the scheme is applicable to
any number of users scheduled simultaneously. To make it possible, we perform superposition coding and
successive interference cancelation (SIC) for the successful transmission of data streams of simultaneously
scheduled users \cite{Ralf1}.

Let ${\mathcal{K}}$ denote the set of users to be scheduled and $\Phi$ be the
permutation of the scheduled user indices that sorts the channel gains in increasing order, i.e.\
$h_{\Phi_1}\le \cdots \le h_{\Phi_k}\le \cdots \le
h_{\Phi_{|{\mathcal K}|}}$. Then, the energy of the scheduled user
$\Phi_k$ with rate $R_{\Phi_k}$, is given by
\cite{Tse3,Ralf1}
\begin{equation}\label{eqn:power}
 E_{\Phi_k} = \frac{Z_0}{h_{\Phi_k}} \left({2^{\sum_{i\leq k}R_{\Phi_i}}-2^{\sum_{i<k}R_{\Phi_i}}}\right)~,
\end{equation}
where $Z_0$ denotes the noise power spectral density.

To ease the understanding of the discussion in the paper, we summarize the notation for the system design
parameters in Table \ref{tab:notation}.
\begin{table}
\centering
\footnotesize
\caption{System Parameters}
\begin{tabular}{lr}
\toprule
Parameter & Symbol\\
\midrule
Buffer Size								& $B$\\
CCON parameter & $N$ \\
Probability of violation of CCON 					& $\gamma$ \\
Target probability of violation of CCON & $\epsilon$\\
Average (target) packet drop rate & $\theta_{\rm tar}$\\
Average packet drop rate achieved as a function\\ of other parameters & $\theta_r$\\
Probability that a transmission is not successful & $\nu_d$\\
Probability to have CCON parameter $N_a$& $\zeta_a$\\
\bottomrule
\end{tabular}
\label{tab:notation}
\end{table}

\subsection{Packet Arrival Model in Large User Limit}
\label{sect:packet_arrival}
The design of the scheme presented later in this work is based on the asymptotic case when the number of the
users approach infinity, i.e., $K\to \infty$. We consider an arbitrary random packet arrival process for a
user $k$ with bounded mean and variance. At the system level, when an asymptotically large number of users
are present, the "system" packet arrival process can be modeled with a constant arrival process
\cite{majid_Eurasip}. Regardless of the arrival distribution, the system level arrival rate converges to
statistical average of the arrival process when an infinitely large number of users are present in the
system. For a single user, this is modeled by the constant arrival of a single packet with variable size in
each time slot where no arrival is modeled by arrival of a packet with zero size\footnote{Zero packet size
facilitates modeling of the scheme (as explained in next section) while arrival (and transmission) of a
packet with zero rate has no effect on system energy consumption.} \cite{majid_TWC:13}.

In the large user limit, multiuser scheduling problem can be broken into a single user scheduling problem
such that every user takes the scheduling decision independent of the other users\footnote{Though, users'
scheduling decisions decouple as a result of large user limit assumption, power allocation for the scheduled
users requires rate information of the other scheduled users. 
information system \cite{majid_TWC:13}.
}\cite{majid_TWC:13}.
The large system results have been employed successfully in communications in different settings to analyze
the systems with dependencies, e.g., \cite{boudec:08,viswanath:01}.

\section{Modeling the Scheduling Scheme}
\label{sect:scheme}
Packet scheduling constrained by average packet drop rate and maximum successive packet drop belongs to a
class of sequential resource allocation problems, known as Restless Multi-armed Bandit Processes (RMBPs)
\cite{Whittle:1988}. In RMBPs, a subset of the competing users are scheduled in each slot. The states of all
the users in the system stochastically evolve based on the current state and the action taken. The scheduled
user receives a reward dependent on its state. The next action depends on the reward received and the
resulting new state. The RMBPs are characterized by a fundamental trade-off between the decisions promising
high immediate rewards versus those that sacrifice immediate rewards for better future rewards. In contrast
to use of RMBPs to model and analyze the effect of correlation between channel states
\cite{Wei:2010,chie:2010}, our optimization problem is based on investigating the effect of sequential
decisions in terms of correlation between packet dropping sequences. The one bit channel feedback does help
to make the decision in the next time slot but it does not give any idea about the channel state in the next
time slot due to block fading model assumption.

The scheduling framework comprises two parts: online scheduling decisions and the offline optimizations of
scheduling thresholds.
The scheduling decisions for every user in each time slot are based on the instantaneous channel condition
and the scheduling thresholds. The thresholds are optimized by taking into consideration the CCON parameter
$N$, maximum buffer size $B$, average packet dropping probability $\theta_{\rm tar}$ and the user's small
scale fading distribution. The number of thresholds equals the number of buffered packets and the scheduler
decides how many packets are scheduled in a single time slot based on the channel conditions. If no packet is
scheduled, all the packets (including the recently arrived packet) are buffered if the buffer has capacity.
If the buffer is full, the oldest packet in the buffer is dropped. When the user has dropped $N$ packets
successively (bursty loss), the scheduling of at least a single packet is maximally prioritized, but it
cannot be guaranteed due to random fading channel. Thus, the lowest scheduling threshold is dependent on the
maximum statistical guarantee $\gamma$ that CCON cannot be fulfilled. $\gamma=0$ is a special case where
scheduling threshold is set to zero when $N$ packets have been dropped successively \cite{majid_TWC:15}.

Next, we address the online scheduling mechanism in Section \ref{sect:FSMC}, while offline threshold
optimization is discussed in Section \ref{sect:optimization}.

\subsection{Finite State Markov Chain Model}
\label{sect:FSMC}
We model the proposed scheduling scheme using a finite state Markov chain (FSMC). Let $i\leq B$ and $j\leq N$
denote the number of
packets buffered and dropped successively at time $t$. Then, the Markov chain state $p$ at time $t$ is
defined by a variable from the composite state space such that $p=i+j$. At
the start of the process, $p$ equals zero. If a packet is not scheduled, it is buffered and $i=1$
(while $j=0$), thereby the system makes transition to next state $q=1$. Remember $p(t+1)=q(t)$ in FSMC. When
the buffer is full, an event of not
scheduling a packet results in a packet drop, thereby $j$ starts increasing and $i=B$ remains fixed until
there is a room in the buffer for unscheduled packets due to scheduling of previously buffered packets. The
event of dropping/buffering of the packet results in a forward state transition to the next state $q=p+1$.
The size of FSMC is determined by the buffer size and CCON parameters such that $M=B+N$.

We consider the event of packet drop due to imperfect CSI in the state space description next. We assume that
feedback for the
successful/unsuccessful transmission (ACK/NACK) arrives by the end of time slot and the transmitter buffers
the scheduled packet(s) by then. If the transmitter receives an ACK, the packets are dropped from the buffer
as they have been received successfully. In case of a NACK, the buffered packets are treated in the same way
as intentional packet dropping, i.e., buffer if there is a room or drop otherwise. The dropping of a packet
in case of a NACK occurs solely due to insufficient buffer capacity and affects system performance similar to
intentional packet drop scenario. The packet drop due to imperfect CSI needs to be modeled in the system
separately due to its different effect on system energy. Intentional packet dropping (without transmission)
does not cost any energy to the system while packets dropped due to imperfect CSI result in waste of energy
without transmitting data successfully.

As explained in Section \ref{sect:packet_level}, the effect of imperfect CSI at the transmitter side is
modeled by packet level description such that $\nu_d$ denotes packet drop probability and $\nu_s=1-\nu_d$ is
the probability of a successful transmission.

Thus, we define state transition probability $\alpha_{pq}$ in an FSMC model as
\begin{eqnarray}
\alpha_{pq} &=& {\rm Pr}(S_{t+1}=q|S_t=p)\\
&=&\begin{cases}
\nu_s\hat{\alpha}_{pq} & p< M,q\leq \min(p,B)\\
\tilde{\alpha}_{pq} + \nu_d\sum_{m=0}^{\min(p,B)}\hat{\alpha}_{pm}&  p< M,q=p+1\\
0& \mbox{else}
\end{cases}
\end{eqnarray}
where
\begin{eqnarray}
\alpha_{pq} &=& \mbox{Transition probability from state $p$ to $q$.}\nonumber\\
\hat{\alpha}_{pq} &=& \mbox{Transition probability from state $p$ to $q$}\nonumber\\&& \mbox{when scheduling
of one or more packets
occurs.}\nonumber\\
\tilde{\alpha}_{pq} &=& \mbox{Transition probability from state $p$ to $q$ when no}\nonumber\\&& \mbox{packet
is scheduled.}\nonumber
\end{eqnarray}
To define $\hat{\alpha}_{pq}$ and $\tilde{\alpha}_{pq}$ mathematically, we define a scheduling threshold.
\begin{definition}[Scheduling Threshold $\kappa_{pq}$] It is defined as the minimum small scale fading value
$f$ required to make a state transition
from state $p$ to $q$ such that
\begin{equation}
\hat{\alpha}_{pq} =
{\rm Pr}\bigl(\kappa_{pq}<f\leq\kappa_{p(q-1)}\bigr)\quad 0\leq q\leq \min(p,B).
\label{eqn:alpha1}
\end{equation}
where $\kappa_{p0^-}$ is defined to be infinity with $S_{0-}$ denoting a dummy state before $S_0$.
\end{definition}
From scheduling point of view, it is advantageous to schedule more packets for good fading states. Therefore,
the scheduling thresholds quantize the
fading vector to optimize the number of scheduled packets according to the fading.

In a state $p\geq q$, the scheduler with fading variable $f$ makes a state transition to state $q$ such that
\cite{majid_TWC:15}
\begin{equation}
q =  \kappa_{p{q}}<f\leq\kappa_{p({q}-1)}\quad 0\leq {q}\leq \min(p,B)~.
\label{eqn:decision}
\end{equation}
For a state transition from state $p$ to $q$, the number of scheduled packets is given by
\begin{equation}
L(p,f) = \min(p,B)-q+1~,
\label{eqn:packets}
\end{equation}
where $q$ is determined uniquely by (\ref{eqn:decision}). Note that the number of scheduled packets cannot
exceed $\min(p,B)$ because of finite
capacity of buffer. We denote $\min(p,B)$ by a variable $\mu=\min(p,B)$ in the rest of this article for
convenience.

The probability of not scheduling any packet for transmission is expressed by
\begin{eqnarray}
\tilde{\alpha}_{pq}&=& F_{f}(\kappa_{p\mu})~,\qquad 0\leq p< M,q=p+1\\
&=&1-\sum_{q=0}^\mu\hat{\alpha}_{pq}
\label{eqn:alpha2}
\end{eqnarray}
where $\kappa_{p\mu}$ denotes the minimum thresholds to schedule at least one packet in state $p$.

\begin{figure}
\centering
   \includegraphics[width=2.0in]{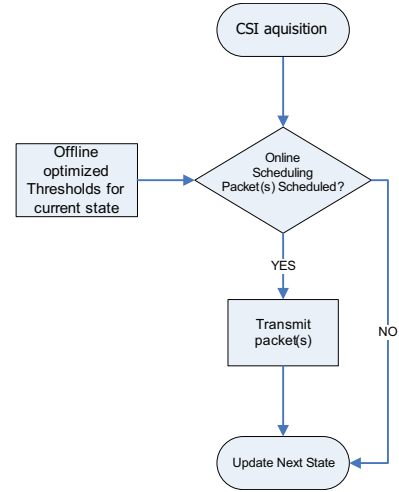}
  \caption{Flow chart for the scheduling mechanism.}
  \label{fig:algo}
 \end{figure}

To further explain the online scheduling mechanism, the flowchart is presented in Fig. \ref{fig:algo}.

\subsection{Modeling $\gamma$ in FSMC}
Ideally one would like to schedule a packet with probability one when $p=M$ and $j=N$. As explained earlier,
this is not practical due to the following constraints:
\begin{enumerate}
  \item It is not possible to apply 'water filling' principle on any arbitrary channel due to power
      limitations of the transmitter. Thus, a packet is not scheduled with probability one in state $M$.
      This is implemented by having $\kappa_{MB}>0$ and not scheduling a packet if $f_k\leq \kappa_{MB}$.
  \item When $\nu_d>0$, it cannot be guaranteed with probability one that the scheduled packets in state
      $M$ are received by the receiver error free.
\end{enumerate}
Both of the constraints contribute to statistical guarantee on CCON with $\gamma>0$.

To handle the event of unscheduled or/and lost head of line (HOL) packet in state $M$, we define a self state
transition $\alpha_{MM}$ where no packet is scheduled in contrast to other self state transitions (where a
single packet is scheduled) with $\tilde{\alpha}_{MM}=\Pr(f_k\leq \kappa_{MB})$.

\begin{figure}
\centering
   \includegraphics[width=3.3in]{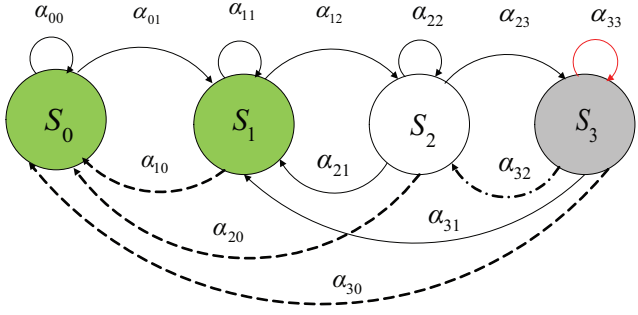}
  \caption{State transition diagram of the scheme for the case $B=2, N=1$. $\alpha_{MM}$ represents state
  transition probability related to $\gamma$.
  }
  \label{fig:state diagram}
\end{figure}

Thus, $\gamma$ is modeled using FSMC model and the constraints above by
\begin{eqnarray}
\gamma&=&\alpha_{MM}\pi_M= \Bigl(\tilde{\alpha}_{MM}+\nu_d \sum_{q=0}^{B}\hat{\alpha}_{Mq}\Bigr)\pi_M\\
&=& \Bigl(1-\nu_s \sum_{q=0}^{B}\hat{\alpha}_{Mq}\Bigr)\pi_M
\label{eqn:outage}
\end{eqnarray}
where $\pi_M$ is steady state transition probability for state $M$.
\begin{example}
Let us explain FSMC model with the help of an example with $B=2$, $N=1$ as in Fig. \ref{fig:state diagram}.
For this example, we evaluate the transition probability matrix $\mat{Q}$.
\end{example}
The steady state transition probability matrix $\mat{Q}$ is expressed as
\begin{equation}
\mat{Q}=\mat{Q_{\rm s}}+\mat{Q_{\rm c}}
\end{equation}
where
\begin{equation}
\mat{Q}_{\rm s} = \left( \begin{array}{cccc}
\nu_s\hat{\alpha}_{00} & \tilde{\alpha}_{01}&0  &0 \\
\nu_s\hat{\alpha}_{10} & \nu_s\hat{\alpha}_{11} & \tilde{\alpha}_{12} &0 \\
\nu_s\hat{\alpha}_{20} & \nu_s\hat{\alpha}_{21} & \nu_s\hat{\alpha}_{22} &\tilde{\alpha}_{23} \\
\nu_s\hat{\alpha}_{30}& \nu_s\hat{\alpha}_{31}& \nu_s\hat{\alpha}_{32} &\tilde{\alpha}_{33}\end{array}
\right) \label{eqn:Drop_STM1}
\end{equation}
and
\begin{equation}
\mat{Q}_{\rm c} = \nu_d \left( \begin{array}{cccc}
0 & \sum_{q=0}^0 \hat{\alpha}_{0q}&0  &0 \\
0 & 0 & \sum_{q=0}^1 \hat{\alpha}_{1q} &0 \\
0 & 0& 0 &\sum_{q=0}^2 \hat{\alpha}_{2q} \\
0 & 0& 0&\sum_{q=0}^2 \hat{\alpha}_{3q}\end{array} \right) \label{eqn:Drop_STM2}.
\end{equation}
$\mat{Q}_{\rm c}$ captures the effect of imperfect CSI while $\mat{Q}_{\rm s}$ is optimized scheduling
decision matrix. Note that this model implies
that it
is not possible to achieve CCON with probability one if $\nu_d>0$ and only statistical guarantees can be
provided with $\gamma>0$.

\section{Mathematical Formulation of the Problem}
\label{sect:optimization}
The objective of the optimization problem is to minimize the system energy for a soft average packet drop
rate constraint and statistical guarantee on CCON. We formulate the optimization problem using the FSMC model
developed in the previous section.
Each scheduled packet is treated as an independent virtual user for the analysis purpose.
For the case of imperfect CSI at the transmitter side, the average system energy per transmitted information
bit at
the large system limit $K \to \infty$ is given by \cite{Ralf1}
\begin{equation}
\Bigl(\frac{E_{b}}{N_0}\Bigr)_{\rm CST}=  \log(2) \int\limits_0^\infty \frac {2^{C \,{\rm P}_{h,\rm
VU}(x)}}x\, {\rm dP}_{h,\rm VU}(x)
\label{eqn:energy_function}
\end{equation}
where ${\rm P}_{h,\rm VU}(\cdot)$ denotes the cdf of the fading of the scheduled virtual users (VUs).

The energy expression in (\ref{eqn:energy_function}) requires channel distribution ${\rm P}_{h,\rm VU}(x)$ of
the scheduled users. In the large system limit, ${\rm P}_{h,\rm VU}(x)$ depends only on the small scale
fading distribution because of the fading-dependent scheduling decisions as the path loss distribution for
the VUs is the same as for the mobile users.
The probability density function (pdf) of the small scale fading of the scheduled VUs is given by
\cite{majid_TWC:15}
\begin{equation}
{\rm p}_{f,\rm VU} (y) = \sum\limits_{p=0}^M c_p\pi_pL(p,y)\,{\rm p}_{f}(y) \label{eqn:SVU_fading}
\end{equation}
where ${\rm p}_{f}(y)$ and $c_p$ denote the
small scale fading distribution and a normalization constant, respectively while $L(p,y)$ is given by
(\ref{eqn:packets}).
The channel distribution for the scheduled VUs is computed using fading distribution in
(\ref{eqn:SVU_fading}) and the
path loss distribution in (\ref{eqn:path_loss}).

Thus, the optimization problem is formulated as
\begin{eqnarray}
\label{eqn:objective}
&\min_{\mat{Q}\in \Omega} \big(\frac{E_{b}}{N_0}\big)_{\rm CST}&\\
&\mbox{s.t.}:\begin{cases}\mathcal{C}_1: 0\leq\sum_{m=0}^{\mu}\alpha_{pm}\leq 1 &   0 \leq \alpha_{pm} \leq
1,\\&0\leq p\leq M\\
\mathcal{C}_2: \theta_r\leq \theta_{\rm tar} &  \mat{Q}\in \Omega\\
\mathcal{C}_3:\sum_{q=0}^M \alpha_{pq}=1& 0\leq p \leq M\\
 \mathcal{C}_4:B+N=M & B<\infty, N<\infty
\end{cases}&
\label{eqn:optimization}
\end{eqnarray}
where $\Omega$ denotes the set of permissible matrices for $\mat{Q}$ and $\theta_r$ is the average packet
drop rate achieved for a fixed $\mat{Q}$ and given by
\begin{eqnarray}
\label{eqn:drop_cons2}
\theta_r &=& \sum_{p=B}^{M-1}\alpha_{p(p+1)}\pi_p+\alpha_{MM}\pi_M\\
&=& \sum_{p=B}^{M}\bigl(1-\nu_s\sum_{m=0}^{B}\hat{\alpha}_{pm}\bigr)\pi_p.
\end{eqnarray}
Equation (\ref{eqn:drop_cons2}) is a result of combining $\mathcal{C}_1$ and $\mathcal{C}_3$ in
(\ref{eqn:optimization}).

The forward transition for the states $B\leq p< M$ and self state transition in state $M$ models the events
of packet drop and the summation
over the probability for these events multiplied by corresponding steady state probabilities results in the
average dropping probability in (\ref{eqn:drop_cons2}). The summation starts from state $B$ as the
unscheduled packets are buffered for $p<B$. For a fixed $p$, the corresponding channel-dependent optimal
scheduling thresholds can be computed from the optimized $\vec{\alpha_p^*}=[\alpha_{p0}^*,\dots
\alpha_{p\mu}^*]$ using (\ref{eqn:alpha1}).
The violation probability on CCON $\gamma$ for fixed $B$ and $N$ parameters is computed from $Q^*$ using
(\ref{eqn:outage}). Let us denote $\gamma$ for this special case by $\gamma_m$ where the \emph{maximum}
energy efficiency can be achieved for fixed $B,N,\theta_{\rm tar}$ parameters and relaxing $\gamma$ further
does not help to improve energy efficiency due to coupling of $\gamma$ with $N$ and $\theta_{\rm tar}$
parameters.

If the statistical guarantees have to be improved further, we apply an upper bound on $\gamma$ such that
$\gamma\leq \epsilon$ where $\epsilon$ is a small constant representing the target statistical guarantee.
This constraint appears as an additional constraint in (\ref{eqn:optimization}) such that
\begin{eqnarray}
      \mathcal{C}_5: \gamma\leq \epsilon,\quad 0\leq\epsilon\leq \theta_{\rm tar}
      \label{eqn:gamma_constraint}
\end{eqnarray}
because $\theta_r = \sum_{p=B}^{M-1} \alpha_{p(p+1)} \pi_p + \gamma$. Consequently, the improved $\gamma$ is
achieved at the increased energy cost. Theoretically $\epsilon$ is upper bounded by $\theta_{\rm tar}$, but
$\gamma_m$ upper bounds $\epsilon$ (tightly) at a
value lower than $\theta_{\rm tar}$ due to the tight coupling of $N$ and $\gamma$ parameters.

It is worth noting that increasing both $N$ or/and $\gamma$ improves energy efficiency. However, the effect
of both parameters on QoE is different. On one side, $N$ bounds the bursty packet loss while on the other
hand, $\gamma$ bounds the events when CCON is violated. By QoE point of view, bounding $\gamma$ is as
critical as bounding $N$ itself and characterizing both is important.

To characterize $\gamma$ as a function of $(N,B,\theta_{\rm tar})$ parameters, we can write $\gamma_m$ as
\begin{eqnarray}
\gamma_m&=&{\rm Pr}(\mbox{More than $M$  packets dropped successively})\nonumber\\
&=& \sum_{a=M+1}^\infty{\rm Pr}(a \mbox{ packets dropped successively})
\end{eqnarray}
As $M=N+B$ where $N$ is a system imposed constraint, we can increase $B$ to reduce the system energy
expenditure. Suppose $\acute{B}=B+1$ and so as $\acute{M}=M+1$. Using above equation, it is clear that
difference in $\gamma_m$ is the probability that exactly $M+1$ packets are dropped successively.
\begin{equation}
\gamma_{B}-\gamma_{\acute{B}}= {\rm Pr}(\acute{M} \mbox{ packets dropped successively})
\end{equation}

\subsection{Trading Buffer for Improved Guarantees on $\gamma$}
\label{subsec:optimizations}
Let us denote $\big(\frac{E_{b}}{N_0}\big)_{\rm CST}$ by $\frac{E_{b}}{N_0}$ for simplicity in rest of this
article. We would like to achieve $\epsilon\leq\gamma_m$ at improved energy by exploiting buffer size as a
degree of freedom and increasing $B$ for a fixed $N$.

Let us denote the optimal solution of the programming problem in previous section by $Q^*(B,\theta_{\rm
tar},\epsilon)$ as a function of $B$, $\theta_{\rm tar}$ and target violation probability on CCON $\epsilon$.
Let $\frac{E_b}{N_0}\big(Q^*(B,\theta_{\rm tar},\epsilon)\big)$ be the corresponding system energy and
$\Delta E$ represents the target energy gain. Now, the optimization is performed
over $B\in \Psi$ where $\Psi$ is a set of possible buffer sizes. For every candidate $B\in\Psi$, optimization
in (\ref{eqn:objective}) and
(\ref{eqn:optimization}) is performed again by including $\mathcal{C}_5$ also. The aim of the optimization is
to find minimum value of $B$ which gives energy less than $\Big(\frac{E_b}{N_0}(Q^*(B,\theta_{\rm
tar},\epsilon))-\Delta E\Big)$ at $\epsilon$:
  \begin{eqnarray}
  	&{\rm Find}~  B^* \in \Phi \quad {\rm s.t.} \quad \gamma \big(Q^*(B^*,\theta_{\rm tar})\big) \leq
  \epsilon \quad {\rm and}&  \\
  	 &\frac{E_b}{N_0}\big(Q^*(B^*,\theta_{\rm tar},\epsilon)\big) - \frac{E_b}{N_0}\big(Q^*(B,\theta_{\rm
  tar},\epsilon)\big) \geq \Delta E ,& ~   B \in \Psi   \nonumber
  \label{eq:form1}
  \end{eqnarray}
The suitable value of $B$ is highly dependent on the application. For example, wireless sensor networks would
prefer large $B$ due to battery
requirements whereas multimedia applications prefer small $B$ due to stringent delay requirements on data
delivery.

\subsection{Stochastic Optimization}
\label{sec:stochastic_optimization}
The optimization problem formulated in (\ref{eqn:objective}) and (\ref{eqn:optimization}) is not convex and
can be solved using
stochastic optimization techniques. There are a few heuristic techniques in literature to solve such problems
like genetic algorithm, Q-learning, neural
networks, etc. We use Simulated Annealing (SA) algorithm to solve the problem. As the name suggests, the
algorithm originates from the statistical
mechanics area and has been quite useful to solve different combinatorial optimization problems like
traveling salesman.

In SA algorithm, a random configuration in terms of transition probability matrix $\mat{Q}$ is presented in
each step and the system energy as an objective
function is evaluated only if $\mat{Q}$ fulfills all the constraints in (\ref{eqn:optimization}). If the
system energy improves the previous best solution,
the candidate configuration is selected as the best available solution. However, a candidate configuration
can be treated as the best solution with a
certain temperature dependent probability even if the new solution is worse than the best known solution.
This step is called \emph{muting} and helps
the system to avoid local minima. The muting step occurs frequently at the start of the process as
temperature is selected very high and decreases as the temperature is decreased gradually. Thus, the term
temperature determines the rate of muting process.

In literature, different cooling temperature
schedules have been employed according to the problem requirements. In this work, we employ the following
cooling schedule, called fast annealing (FA) \cite{FA}.
In FA, it is sufficient to decrease the temperature linearly in each step $b$ such that,
\begin{equation}
\label{eqn:FA} T_b = \frac{T_0}{c_{\rm sa}*b+1}
\end{equation}
where $T_0$ is a suitable starting temperature and $c_{\rm sa}$ is a constant which depends on the problem
requirements. The parameters of the temperature schedule can be computed via experimentation, e.g., as in
\cite{majid_TWC:13,Samer:2012}.
The pseudocode for the optimization of programming problem using SA is presented in Algorithm
\ref{algorithm}.

\begin{algorithm}
\caption{Optimization by SA Algorithm}
\KwIn{$(\mat{Q_0}, T_m,\theta_{\rm tar},\epsilon$)\;}
$E_0$= Compute energy as a function of initial $\mat{Q_0}$\;
$E^* =E_0$;$\mat{Q^*} = \mat{Q_0}$\;
$T=$ New lower temperature according to FA schedule\;
\tcc{Perform temperature iterations as long as it reaches the lowest temperature $T_m$.}
\While{T $\geq T_m$}{
\tcc{Generation of $n$ random configurations for temperature $T$.}
\For{i=0 \KwTo n}{
 Generate a random $\mat{\hat{Q}}$\;
 Compute $\gamma$ and $\theta_r$ for $\mat{\hat{Q}}$\;
 \tcc{Evaluating $\mathcal{C}_2$ and $\mathcal{C}_5$.}
  \If{$(\theta_r<\theta_{\rm tar}\AND\gamma\leq \epsilon$)}{
  Compute energy $\hat{E}$ as a function of $\mat{\hat{Q}}$\;
  $r$ = A random number in range $[0,1]$\;
  \tcc{Implementation of Muting step.}
  \If {$r<\exp \big(\frac{-(\hat{E}- E^*)}{T}\big)$}{
        $\mat{Q^*}=\mat{\hat{Q}}$\;
        \tcc{Energy update step.}
     \If {($\hat{E}\leq E^*$)}{
        $E^* =\hat{E}$;
        }
     }
   }
  }
}
\KwOut{$(E^*,\mat{Q^*}$);}
\label{algorithm}
\end{algorithm}


\subsection{Physical Layer Channel Estimation Model}
In contrast to packet level channel model for imperfect CSI at the transmitter side, the effect of imperfect
CSI at both transmitter and the receiver sides is modeled at physical channel level by a channel estimation
error variance. The receiver performs pilot (or data) aided channel estimation by some criterion, e.g.,
Linear Minimum Mean Square Error (LMMSE). The resulting error in estimation is modeled by certain variance
$\beta^2$ that depends on the pilot signal length and power. Note that there is no feedback channel available
and the user does not adapt his scheduling decision if a transmitted packet is dropped. In fact, physical
level channel model is oblivious of the packet level scheduling and determines bit level performance.

The channel estimation error results in higher energy per bit. The average system energy per transmitted bit
for this case is derived in Appendix \ref{app:imperfectCSI} and given by
\begin{eqnarray}
\label{eqn:energy_error}
	\Big(\frac{E_{b}}{N_0}\Big)_{\rm CSO}  & = & \log(2) \int_{0}^{\infty} \frac{2^{C {\rm P}_{h,\rm
VU}(x)}} {x} {\rm dP}_{h,\rm
VU}(x) \label{eq:rf} \\
	& & + \beta^2 \log(2) \int_0^\infty \frac{2^{2C {\rm P}_{h,\rm
VU}(x)}} {x^2}{\rm dP}_{h,\rm
VU}(x) \nonumber.
\end{eqnarray}
Regardless of the scheduling scheme at link layer, the transmit power can be adapted as a function of error
variance $\beta^2$. To eliminate the effect of channel estimation error, the user transmits with an extra
power margin where margin is calculated as a function of $\beta^2$ such that the effect of estimation error
can be removed. We model this scenario by considering $\nu_d=0$ (error free transmission) in our scheduling
scheme such that the transmission requires $\big(\frac{E_{b}}{N_0}\big)_{\rm CSO}$ instead of
$\big(\frac{E_{b}}{N_0}\big)_{\rm CST}$ for the same system parameters.

\section{Modeling Individual User CCON Constraints}
\label{sect:generalization}
We generalize our framework to the case when the individual users have non-identical CCON parameter $N_a$,
where $a\in \{1,2,\dots A\}$. To model the general case at system level, we define a system level CCON
parameter $N$ by,
\begin{equation}
N=\max\{N_1,N_2\dots N_A \}.
\end{equation}
We denote the probability that a user has a CCON parameter $N_a$ by $\zeta_a\geq 0$ such that $\sum_{a=1}^A
\zeta_a=1$. Note that $\zeta_a$ can be zero for some $N_a$.

\begin{figure}[t]
 \centering
 \includegraphics[width=3.1in]{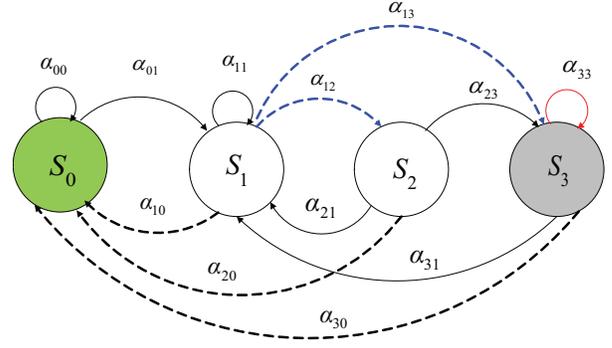}
  \caption{The Markov chain model for a system with $B=1$ and the users have CCON parameters $1$ and $2$ with
  probability $\zeta_1$ and $\zeta_2$, respectively. The system level state diagram shows the modeling at
  system level where $N=\max\{1,2\}$.
  }
  \label{fig:system_level}
\end{figure}

To explain the concept, let us discuss the example when the users have CCON parameters $1$ and $2$ such that
$\zeta_1$ and $\zeta_2$ proportion of the users have the constraint $1$ and $2$, respectively. Buffer size is
fixed to one for both cases. In contrast to the case with homogenous $N$, the system level Markov chain will
be different from the user level model. The individual users will have state space model corresponding to
$B=1,N_a=1$ and $B=1,N_a=2$ cases (as modeled before), but the resulting (cumulative) system space model is
shown in state diagram in Fig.~\ref{fig:system_level} such that $N=\max\{1,2\}$ and $\zeta_1$ and $\zeta_2$
denote the respective probabilities of having $N_1$ and $N_2$.

For the individual CCON parameter case, the state transition probabilities and the resulting steady state
probabilities are modified. For example,
\begin{eqnarray}
\pi_{2}&=& \zeta_2\big(1-\nu_s(\hat{\alpha}_{10}+\hat{\alpha}_{11})\big)\pi_{1}\\
\pi_{3}&=&\zeta_1\big(1-\nu_s(\hat{\alpha}_{10}+\hat{\alpha}_{11})\big) \pi_1+\alpha_{23}\pi_2
\end{eqnarray}
where the state transition probabilities are calculated in the same way as in Section
\ref{sect:optimization}.

In general,
\begin{eqnarray}
\pi_{p}&=& \zeta_N\big(1-\nu_s\sum_{q=0}^B\hat{\alpha}_{Bq}\big)\pi_{B},\quad p=B+1 \\
\pi_{p}&=& \alpha_{(p-1)p}\pi_{p-1}+\alpha_{Bp}\pi_B, \quad  B+1<p\leq M\\
&=&\big(1-\nu_s\sum_{q=0}^B\hat{\alpha}_{pq}\big)\pi_{p-1}+\zeta_{M-p+1}\big(1-\nu_s\sum_{q=0}^B\hat{\alpha}_{Bq}\big)\pi_B\nonumber
\end{eqnarray}
while the steady state probabilities for the states $p\leq B$ do not depend on the distribution of $N$ and
calculated as before.

Similarly, the average packet drop rate in (\ref{eqn:drop_cons2}) is modified as,
\begin{eqnarray}
\theta_r = \sum_{p=B}^{M-1}\alpha_{p(p+1)}\pi_{p}+\alpha_{MM}\pi_M+\pi_B\sum_{q=B+2}^M {\alpha}_{Bq}.
\label{eqn:drop_cons_general}
\end{eqnarray}
After some mathematical manipulation, it can be shown that
\begin{eqnarray}
\sum_{q=B+2}^M {\alpha}_{Bq}&=& \sum_{q=B+2}^M \zeta_{M-q+1}\tilde{\alpha}_{Bq}\\
&=&(1-\zeta_N)\big(1-\nu_s\sum_{q=0}^B\hat{\alpha}_{Bq}\big).
\end{eqnarray}
The additional term represents the packets dropped as a result of having $N_a<N$.
It is worth noting that $\theta_r$ is the system level parameter and an upper bound on $\theta_r$ for the
individual users.
The individual users with $N_a<N$ may not able to fully utilize it completely for achieving energy efficiency
as average energy saturates at lower $\theta_{r}$ for small values of $N_a$ parameter \cite{majid_TWC:15}.

It is clear from the system state space model that the probability distribution of CCON parameter affects the
system energy efficiency. If the probability of having small $N_a$ is high as compared to the large $N_a$,
the average system energy increases.
We evaluate the effect of this distribution on system energy through numerical simulations in Section
\ref{sect:results}.

\section{Numerical Results and Discussion}
\label{sect:results}
\begin{figure}
\centering
   \includegraphics[width=3.5in]{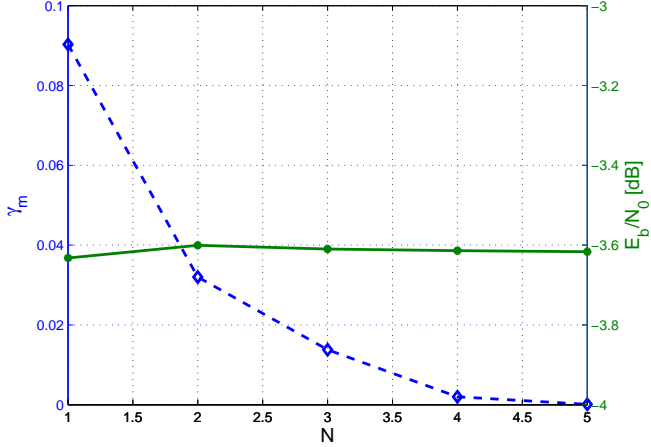}
  \caption{$\gamma_m$ and $\frac{E_b}{N_0}$ as a function of $N$ for our proposed scheme. To better visualize
  the behaviour of $\gamma_m$ and $\frac{E_b}{N_0}$ against $N$ (on x-axis) simultaneously, we plot both
  parameters in the same figure such that the color of the curve for a parameter matches with the color of
  the corresponding y-axis. $B$ is fixed to zero while $\theta_{\rm
tar}=0.3$ and $\nu_d$ equals $0.02$.}
  \label{fig:gamma_lim}
 \end{figure}

We assume that the users are placed uniformly at random
in a circular cell except for a forbidden region around the access point of
radius $\delta=0.01$ according to path loss model in (\ref{eqn:path_loss}). The path loss exponent equals 2
and the
path loss distribution follows the model in \cite{Ralf1}. All the users experience independent small-scale
fading with
exponential distribution with mean one. Spectral efficiency is $0.5$ bits/s/Hz for all simulations. In SA
algorithm, $100$ temperature values are simulated according to FA temperature schedule while $50(M+1)$ random
configuration of transition probability matrix are generated for a single
iteration at temperature $T_b$. The cooling schedule parameters in (\ref{eqn:FA}) are computed after
extensive experimentation such that muting occurs frequently at high temperature and almost seizes at low
temperature.

\begin{figure}
\centering
  \subfigure[System energy as a function of $\epsilon$ for a system with fixed
  $B=0$.]{\includegraphics[width=3.5in]{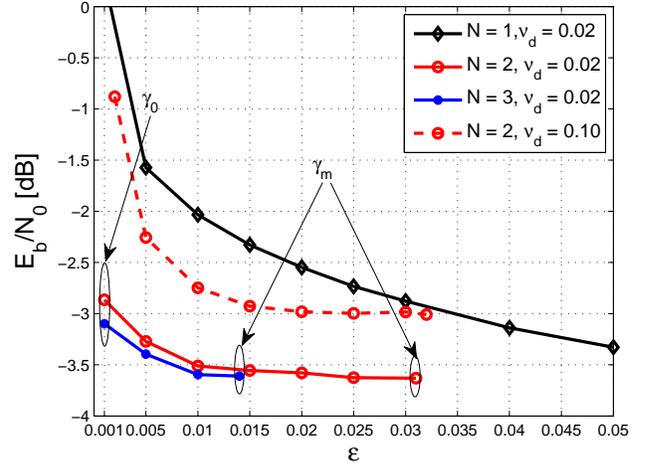}
 \label{fig:energy_nobuffer}}
 \subfigure[Achieved packet drop rate $\theta_r$ from (\ref{eqn:drop_cons2}) as a function of $\epsilon$ for
 the same parameters as in Fig.
 \ref{fig:energy_nobuffer}.]{\includegraphics[width=3.5in]{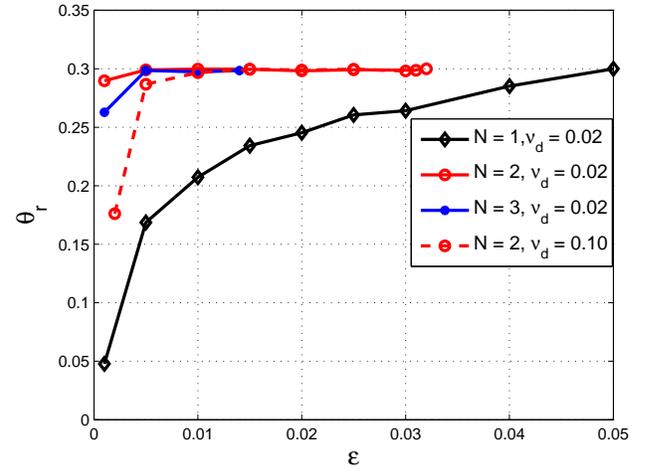}
  \label{fig:drop_nobuffer}}
 \caption{System energy and packet drop behavior as a function of $\epsilon$. $\theta_{\rm tar}$ is fixed to
 $0.3$ for all simulations.}
 \label{fig:energy_vs_gamma}
\end{figure}

Fig.~\ref{fig:gamma_lim} illustrates $\gamma_m$ values and the corresponding system energy (plotted against
right side y-axis) for different $N$ and fixed $B=0$. To compute $\gamma_m$, we perform optimization in
(\ref{eqn:optimization}) without applying constraint in
(\ref{eqn:gamma_constraint}) and the \emph{best}\footnote{We avoid using term energy optimal here as SA is a
heuristic algorithm and solution cannot be proven optimal.} solution matrix $Q^*$ is obtained. The value of
$\gamma$ computed via (\ref{eqn:outage}) for $Q^*$ gives us $\gamma_m$ and upper bounds $\epsilon$.
$\gamma_m$ decreases exponentially with increasing $N$ and reaches nearly zero for $N=5$ while $E_b/N_0$
remains constant for every $(N,\gamma_m)$ tuple. Although, energy per bit for any two different
$(N,\gamma_m)$ pairs is the same, their effect on QoE may vary considerably and dictates which parameter
needs to be employed. Based on numerical results in Fig.~\ref{fig:gamma_lim}, we evaluate the tradeoffs
addressed in Section \ref{subsec:optimizations}.

Fig.~\ref{fig:energy_vs_gamma} exhibits the effect of imposing constraint $\epsilon\leq \gamma_m$ on system
performance when $\theta_{\rm tar}=0.3$. We evaluate $\mathcal{C}_5$ alongwith $\mathcal{C}_1-\mathcal{C}_4$
in (\ref{eqn:optimization}) for the candidate $\mat{Q}$ before evaluation of (\ref{eqn:energy_function}) in
SA algorithm. We observe in Fig.~\ref{fig:energy_nobuffer} that decreasing $\epsilon$ has an associated
energy cost and the solution becomes suboptimal by energy point of view.
Moreover, $\gamma$ can never approach zero as long as $\nu_d>0$ and packet dropping due to imperfect CSI
cannot be completely eliminated. For a given
set of parameters and fixed $\nu_d$, the minimum value of achievable $\epsilon$ is denoted by $\gamma_0$
which lower bounds $\epsilon$ such that
$\gamma_0\leq\epsilon\leq\gamma_m$. The greater the value of $\nu_d$, the greater is $\gamma_0$. For
instance, increasing $\nu_d$ from $0.02$ to $0.1$ for the case $N=2$ raises $\gamma_m$ from $0.001$ to
$0.002$ while system energy increases for all values of $\epsilon$ as well. We observe that bounds on
$\epsilon$ (in the form of $\gamma_0$ and $\gamma_m$) become tight as $N$ increases for the fixed
$\theta_{\rm tar}$. This is due to the fact that allowing large $N$ increases degrees of freedom for the
system and the effect of parameter $\epsilon$ on system energy is minimized.

Correspondingly, Fig.~\ref{fig:drop_nobuffer} demonstrates that achieved average packet drop rate $\theta_r$
(calculated via (\ref{eqn:drop_cons2}))
approaches $\theta_{\rm tar}$ for large $\epsilon$ and remains almost identical thereafter. This implies that
all the extra energy cost is contributed
by strict statistical guarantees on CCON. When $\epsilon$ is very small, the energy optimal $Q^*$ provides a
$\theta_r$ which is much less that
$\theta_{\rm tar}$ and severely sub optimal. We conclude that a strict statistical guarantee on CCON has a
severe plenty in terms of energy and even other DoF (like relaxed $\theta_{\rm tar}$) cannot be utilized
efficiently.

\begin{figure}
\centering
   \includegraphics[width=3.5in]{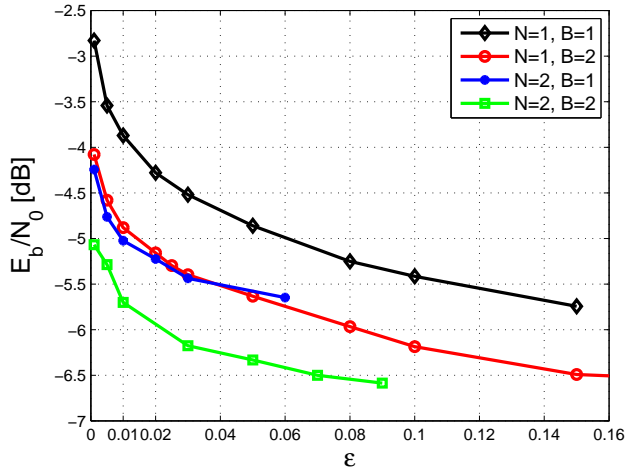}
  \caption{System energy as a function of $\epsilon$ when $B>0$.}
  \label{fig:energy_buffer}
\end{figure}

Fig.~\ref{fig:energy_buffer} demonstrates the energy benefit achieved by increasing buffer size as described
in Section~\ref{subsec:optimizations}. First, we observe that increasing the value of $B$ for a fixed $N$
increases $\gamma_m$, i.e., more flexibility in $\epsilon$.
Secondly, an energy gain by increasing $B$ for all $\epsilon$ and a fixed $N$ is evident. It depends on the
system design that which $B$ needs to be employed for a particular performance guarantee. Let us discuss the
case for parameters $N=1,\theta_{\rm tar}=0.3,\epsilon=0.01$. The system with $B=0$ provides system energy of
almost $-2$ dB as shown in Fig.~\ref{fig:energy_nobuffer}. If we want the same performance at reduced energy,
$B=1$ provides a gain of $\Delta E=1.9$ dB. If $\Delta E>1.9$ dB is desired, $B>1$ is required. For the same
set of parameters, $B=2$ provides $\Delta E$ equal to 3.1 dB. A similar comparison can be drawn for $N=2$ and
$B>0$.

A comparison of the curves for the cases $N=2,B=1$ and $N=1,B=2$ (with same $M=3$) shows that increasing DOF
in any parameter $(B,N)$ is energy efficient as compared to the case $N=1,B=1$, but the effect differs widely
in many ways, e.g., value of $\gamma_m$ for both cases. Similarly, increasing $B$ to reduce system energy
affects system cost while increasing $N$ costs performance loss in terms of jitter.
Thus, system's energy, packet loss and latency requirements determine the parameters required to achieve
performance in terms of statistical guarantee on CCON.
\begin{figure}
\centering
   \includegraphics[width=3.5in]{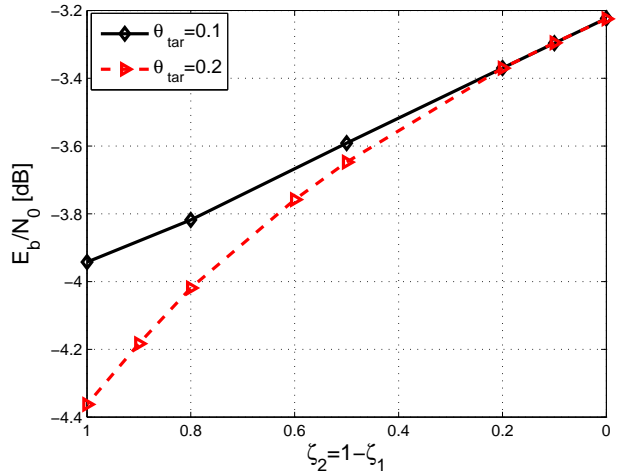}
  \caption{System energy as a function of CCON parameter distribution. Buffer size is fixed to one. To focus
  on the effect of CCON distribution, we set $\nu_s=1$ and $\gamma=0$ while CCON parameters are $1$ and $2$
  with probability $\zeta_2=1-\zeta_1$.}
  \label{fig:energy_nonid}
 \end{figure}

In Fig. \ref{fig:energy_nonid}, we evaluate the effect of CCON parameter distribution on system energy. We
confine ourselves to the case of CCON parameters $1$ and $2$  with probability $\zeta_2=1-\zeta_1$. We see
that system energy decreases as $\zeta_2$ increases. Note that $\zeta_2=0$ implies that all the users have
CCON parameter $1$ while large $\zeta_2$ implies more users with CCON parameter $2$ and more DoF in energy
efficient packet scheduling.

 \begin{figure}
\centering
   \includegraphics[width=3.5in]{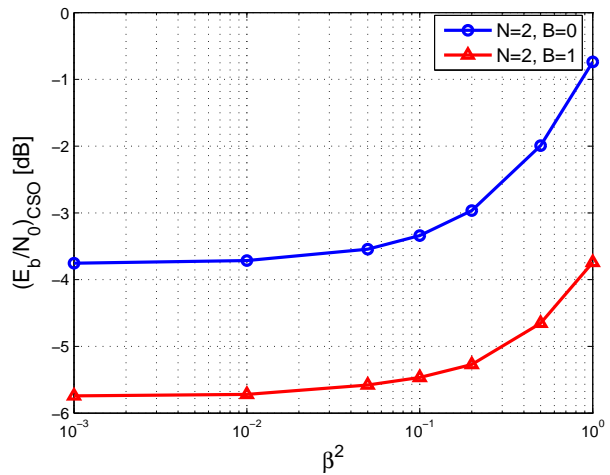}
  \caption{System energy as a function of estimation error variance $\beta^2$. The parameter $\epsilon=0.3$
  while $\gamma=\gamma_m$.}
  \label{fig:energy_var}
 \end{figure}

We demonstrate the effect of estimation error variance on system energy in Fig.~\ref{fig:energy_var} when
imperfect CSI at both transmitter and receiver sides is modeled at physical layer level. We assume that the
effect of error variance remains fixed for all (high and low) signal to noise ratios. As error variance
increases, $\big(\frac{E_{b}}{N_0}\big)_{\rm CSO}$ increases correspondingly. Note that error variance model
does not capture packet level performance and its effect on other packet dropping parameters cannot be
determined. The scheduling decisions are adapted in packet level model as a function of packet loss
probability $\nu_d$ and therefore, both $\big(\frac{E_{b}}{N_0}\big)_{\rm CST}$ and the packet drop design
parameters, (e.g., $\gamma_m,\gamma_0$) change as well. In a physical layer model, no adaptive action is
taken by the scheduler and only $\big(\frac{E_{b}}{N_0}\big)_{\rm CSO}$ is affected by error variance. The
effect of estimation error can be eliminated by transmitting with an extra power margin. It can be observed
from Fig.~\ref{fig:energy_var} that the power margin is high if error variance is large and the increase is
exponential.
\subsection{Discussion}
One of the key features of 5G wireless networks is the availability of services with highly variable QoS
parameters in terms of delay and loss requirements. This work establishes a framework where individual
demands on QoS of the end users are satisfied and energy is saved by exploiting the relaxation in service
guarantees. We deal with the scenarios with erroneous CSI and limited feedback, which reduce the control
traffic significantly.

The analysis of the framework is based on the case with large number of users in the system, which helps to
decouple the scheduling decisions. This implies that the scheme does not suffer from scalability issues, and
actually benefits from more users. However, due to superposition coding, a central unit is required for
sharing CSI information \cite{majid_TWC:15}. As the user threshold optimization is based on the channel
distribution and not the actual realization, the users perform optimization offline and make simple
comparison of thresholds with the available channel state realization to make the scheduling decisions. Thus,
the complexity of the online user scheduling decisions is very small.

\section{Conclusions}
\label{sect:conclusions}
We address the problem of energy efficient multiuser scheduling over fading channels for the loss tolerant
applicants. The packet loss tolerance is characterized by different parameters controlling the QoE for a
specific application. A cross layer framework is proposed and an optimization problem is formulated with the
goal to minimize system energy such that application loss tolerance parameters are satisfied while scheduling
is performed opportunistically over fading channels. We model the framework using FSMC and solve the
optimization problem using simulated annealing optimization technique. We consider the effect of channel
uncertainties on the performance using both channel and packet level methods. Then, the framework is
generalized to the case when bursty packet drop protection varies with the users and model its effect at the
system level.

The results demonstrate the system energy as a function of loss tolerance parameters. We show that buffer
size can be treated as a degree of freedom to improve the QoE for the application constrained by loss
tolerance bounds. An increase in buffer size from one to 2 helps to reduce energy by almost one dB for the
same $\epsilon$. As loss tolerance parameters are coupled, it is not possible to achieve energy efficiency
beyond certain limits by relaxing other parameters if one of the bounds is tight. We conclude that it is
important to exploit DoF available through application loss tolerance to maximize the energy efficiency, and
it is equally important to determine the practical limits on all the parameters which control QoE of the
applications.

\appendices
\section{Derivation of $\big(\frac{E_{b}}{N_0}\big)_{\rm CSO}$ for Imperfect Transmitter and Receiver CSI}
\label{app:imperfectCSI}
In \cite{Medard2000}, a lower bound on the achievable rate region for a two-user MAC with imperfect CSI is
derived. Let us denote the channel estimation error variance as $\beta^2$ and the channel gains by $h_k$ for
$k=1...K$. For $K$ users with fixed power allocation, the achievable rate region is characterized in
\cite[Section III.B]{Medard2000} for all subsets $\mathcal{S} \subseteq \{1,...,K\}$ by
\begin{eqnarray}
	\sum_{k \in \mathcal{S}} R_k \leq \frac{1}{2} \log \left( 1 + \frac{ \sum_{k \in \mathcal{S}} h_k
E_k}{Z_0 + \beta^2 \sum_{k \in \{1,...,K\}}
E_k } \right) \label{eq:ar}.
\end{eqnarray}
Similar to \cite{Tse2}, it can be shown that the minimum energy for fixed rate requirements is achieved for a
decoding order in which the
channel gains $h_1,...,h_K$ are sorted in increasing order. The corresponding power region for the fixed
rates $R_1,...,R_K \geq 0$ is given by the solution of the following linear system of equations
\begin{eqnarray}
	\mat{E}^* = Z_0 \left[ \beta^2 \mat{R} + \mat{B} \right]^{-1} \mat{\rho},
	\label{eq:ezs}
\end{eqnarray}
with rate allocation vector $\mat{\rho} = [\rho_1,...,\rho_K]$ and $\rho_k = 2^{R_k}-1$, coupling matrix
$\mat{B}$
\begin{eqnarray}
	\mat{B} = \left( \begin{matrix} h_1 & - \rho_1 h_2 & - \rho_1 h_3 & ... & - \rho_1 h_K \\
	0 & h_2 & - \rho_2 h_3 & ... & - \rho_2 h_K \\
	\vdots & & \ddots & & \vdots \\
	0 & ... & 0 & 0 & h_K \end{matrix} \right),
\end{eqnarray}
and rate matrix $\mat{R}$
\begin{eqnarray}
	\mat{R} = \left( \begin{matrix} \rho_1 & \rho_1 & ... & \rho_1 \\
	\rho_2 & \rho_2 & ... & \rho_K \\
	\vdots & & \ddots & \vdots \\
	\rho_K & \rho_K & ... & \rho_K \end{matrix} \right).
\end{eqnarray}	
For perfect CSI, i.e., $\beta=0$, the corresponding required transmit power is given by
(\ref{eqn:energy_function}). Let us denote the required transmit power as a function of the channel
estimation error by $\mat{E}^*(\beta)$. For perfect CSI, the transmit power in (\ref{eqn:energy_function}) is
given by $\mat{E}^*(0)$.

Since $\mat{R}$ is rank one, the transmit power in (\ref{eq:ezs}) can be rewritten as
\begin{eqnarray}
	\mat{E}^*(\beta) & = & Z_0 \mat{B}^{-1} \mat{\rho} + \frac{ Z_0 \beta^2 \mat{B}^{-1} \mat{R}
\mat{B}^{-1}}{1 - \beta^2 \text{tr} (\mat{R}
\mat{B}^{-1})} \mat{\rho} \nonumber \\
	& = & \mat{E}^*(0) + \frac{ Z_0 \beta^2 \mat{B}^{-1} \mat{R} \mat{B}^{-1}}{1 - \beta^2 \text{tr} (\mat{R}
\mat{B}^{-1})} \mat{\rho}
\label{eq:r1}.
\end{eqnarray}
This clearly shows the additional power required for imperfect CSI. In order to approximate the second
additional term in (\ref{eq:r1}), we apply the
approximation $\mat{A} \mat{B} \mat{A} \mat{x} \approx 1/K \cdot \text{tr} (\mat{A} \mat{B}) \mat{A}
\mat{x}$. The required transmit power reads
\begin{eqnarray}
	\mat{E}^*(\beta) \approx \left( 1 + \frac{1}{K} \frac{x}{1-x} \right) \mat{E}^*(0) \label{eq:r2}.
\end{eqnarray}	
For small estimation errors, only the first order term of the Taylor series of $\frac{1}{1-x}$ is kept and we
obtain the approximation
\begin{eqnarray}
	\mat{E}^*(\beta) \approx \left( 1 + \frac{1}{K} \beta^2 \text{tr} (\mat{R} \mat{B}^{-1}) \right)
\mat{E}^*(0) \label{eq:r3}.
\end{eqnarray}
The trace can be directly evaluated as $\text{tr} (\mat{R} \mat{B}^{-1}) = \sum_{k=1}^K \frac{\rho_k}{h_k}$.
Using the partial rates (as in
\cite{Ralf1})
$R_k = \lambda_k\frac{C}{K}$, we obtain
\begin{eqnarray}
	\mat{E}^*(\beta)&= & \left( 1 + \beta^2 \frac{1}{K} \sum_{k=1}^K \frac{\exp \left[ \lambda_k\frac{C}{K}
\right] }{h_k} \right) \frac{1}{C} \label{eq:r4}
\\ & & \times  \sum_{k=1}^K \frac{1}{h_k} \exp \left[ \frac{C}{K} \sum_{i < k} \lambda_i \right] \left( \exp
\left[ \lambda_k\frac{C}{K} \right] -
1\right). \nonumber
\end{eqnarray}
For large $K$, $\exp \left[ \lambda_k\frac{C}{K} \right] \approx \lambda_k\frac{C}{K}$ and using \cite[Lemma
1]{Ralf1}, we derive the limiting representation of $\big(\frac{E_{b}}{N_0}\big)_{\rm CSO}$ in
(\ref{eqn:energy_error}).

\bibliographystyle{IEEEtran}
\bibliography{bibliography}

\end{document}